# Variational computation of a vibrating lithium niobate rectangular plate

O. I. Polovina, V. V. Kurylyuk, and O. A. Korotchenkov

*Department of Physics, Kiev National University, Kiev 03680, Ukraine*

**ABSTRACT**

Electrically-excited electroelastic extensional vibrations of an arbitrarily cut three-dimensional piezocrystal resonator are analyzed variationally. A set of trigonometric trial functions applicable to a waveguide behavior of the resonator partly covered by metal electrodes is proposed. The dependence of the content of the vibration modes sustained by the resonator on the electrode configuration is found. The frequency spectra taken in $128^\circ$-Y-rotated $LiNbO_3$ rectangular plates exhibit a good correspondence with the computed resonator eigenfrequencies.

## I. INTRODUCTION

Since a report of Rocke *et al.*[1] the dynamical behavior of photoexcited electrons and holes in semiconductor nanostructures modulated by acoustic waves is a topic of increasing interest. The interaction between the free charges and the waves is mostly due to a piezoelectric coupling. In many cases, the size of the acoustoelectric interaction is not large enough because of a weak piezoelectricity of the structures. A greatly enhanced acoustoelectric interaction can be achieved in a hybrid structure which contains a $LiNbO_3$ plate and a semiconductor nanostructure mounted on it. In these structures, the moving piezoelectric potential of the surface acoustic waves launched in the $LiNbO_3$ plate and the acoustoelectric interaction between the wave and free charges placed in close proximity to the plate surface have been extensively addressed both theoretically and experimentally.[2] The piezoelectric fields of the Lamb waves in $LiNbO_3$ plates has also been reported.[3] Meanwhile, interesting consequences are expected to arise from the interaction of the standing-wave piezoelectric field excited in the $LiNbO_3$ plate and the two-dimensional electron and hole gases created in the adjacent semiconductor nanostructure.

An important advantage of a rectangular plate geometry for the hybrid structure is a possibility to utilize it in a broad frequency range operating either at the standing-wave resonances or at its natural resonant frequencies, thus providing more feasibility in choosing both electrical and elastic field-shapes. Therefore, the major driving force for the study of the piezoelectric-semiconductor structure is the technological importance of the acoustoelectric interaction in the fabrication of nanoelectronic devices. However, one of the limiting restrictions for a broader utilization of these structures is the problem of an accurate theoretical treatment of a vibrating rectangular piezocrystal plate which is partly covered by exciting electrodes.

It is well known that accurate modeling of three-dimensional (3D) electroelastic vibrations can be quite complex in the most general case and requires the application of numerical methods. Variational,[4-6] finite-element,[7,8] and some other[9,10] approximate methods are therefore widely used. The systematic studies have however been only done in piezoceramic circular disks.[11-15] Several interesting issues related to 3D electroelastic vibrations in rectangular-shaped piezoelectric plates,[16-17] piezoelectric parallelepipeds,[4,18,19] and piezo-elastic laminae[20-22] have also been raised. Given the part of the piezoelectric material boundary covered with electrodes is much less than the uncovered part of the boundary, the transfer function, the group delay time, and the impedance matrix have been computed in a rectangular piezoceramic plate sustaining the lowest-order symmetrical Lamb-wave by applying the finite-element method.[17] The present study develops a new approach to the variational treatment of a 3D $LiNbO_3$ rectangular-plate resonator by taking the waveguide behavior of the resonator partly covered by electrodes into account. The electrode configurations providing the excitation of certain vibration modes in the plate are furthermore discussed.

**II. FORMULATION OF THE ELECTROELASTIC INTERACTION FUNCTIONAL**

The general variational formulation of electroelasticity in bounded piezodielectric bodies, based on the Hamilton's variational principle,[23] was elaborated by Holland and EerNisse.[4,5] The method involves the Rayleigh-Ritz calculation procedure and has been successfully used for studying vibrations of a piezoceramic parallelepiped with short-circuited electrodes[4] and a two-electroded piezoceramic disk.[5]

Figure 1(a) shows a schematic sketch of the piezocrystal rectangular plate used in the formulation of the electroelastic interaction functional. Figure 1(b) displays a model geometry used in this work for numerical computations. The electroelastic vibration of the plate is excited by applying a set of $Q_E$ ac-signals to $Q_E$ exciting electrodes shown by filled strips in Fig.1(a), where the *p*-th electrode with the area of $S(p)$ is fed by the signal with the amplitude of $V_G(p)$ and the angular frequency of $\omega$. The plate vibrations are affected by a system of $Q_F$ electrically unloaded, or "free", electrodes with the area of $\Sigma(q)$ [open strips in Fig. 1(a)], due to the screening of the piezoelectric fields by the electrodes. The electrical potential $V(q)$ developed on the *q*-th free electrode is to be determined.

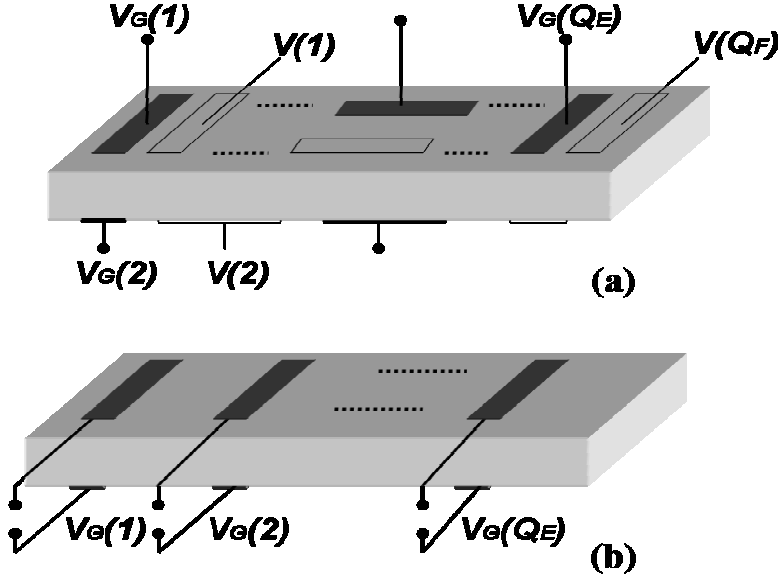

**Fig. 1.** Schematic sketch of a sample for formulation of the interaction functional (a) and model geometry for numerical computations (b). The sample consists of a LiNbO$_3$ plate (shaded body) and a set of exciting (filled strips) and free (open strips) electrodes. The total number of exciting and free electrodes are $Q_E$ and $Q_F$, respectively. They are assumed to be thin enough causing no mechanical load on the surfaces of the plate. Note the absence of the free electrodes on the surfaces and a bulk excitation of plate vibrations in (b).

The functional $F$ describing electroelastic vibrations in the configuration of Fig.1(a) may be expressed as[4,5]

$$F(U_i;\Phi;V(q)) = \frac{1}{2}\iiint_{\Omega_R} g_{i\alpha}g_{j\beta}g_{k\gamma}g_{l\delta}c^E_{\alpha\beta\gamma\delta}U_{i,j}U_{k,l}d\Omega - \frac{1}{2}\omega^2\iiint_{\Omega_R}\rho U_i U_i d\Omega +$$

$$+ \iiint_{\Omega_R} g_{m\alpha}g_{k\beta}g_{l\gamma}e_{\alpha\beta\gamma}\Phi_{,m}U_{k,l}d\Omega - \frac{1}{2}\iiint_{\Omega_R} g_{m\alpha}g_{n\beta}\varepsilon^S_{\alpha\beta}\Phi_{,m}\Phi_{,n}d\Omega -$$

$$- \iiint_{\Omega_E} g_{m\alpha}g_{k\beta}g_{l\gamma}e_{\alpha\beta\gamma}E_{Gm}U_{k,l}d\Omega + \frac{1}{2}\iiint_{\Omega_E} g_{m\alpha}g_{n\beta}\varepsilon^S_{\alpha\beta}E_{Gm}\Phi_{,n}d\Omega -$$

$$- \frac{1}{2}\iiint_{\Omega_E} g_{m\alpha}g_{n\beta}\varepsilon^S_{\alpha\beta}E_{Gm}E_{Gn}d\Omega +$$

$$+ \sum_{p=1}^{Q_E}\iint_{S(p)} N_m(p)g_{m\alpha}g_{k\beta}g_{l\gamma}e_{\alpha\beta\gamma}U_{k,l}[\Phi - V_G(p)]dS -$$

$$- \sum_{p=1}^{Q_E}\iint_{S(p)} N_m(p)g_{m\alpha}g_{m\beta}\varepsilon^S_{\alpha\beta}\Phi_{,n}[\Phi - V_G(p)]dS +$$

$$+ \sum_{q=1}^{Q_F}\iint_{\Sigma(q)} N_m(q)g_{m\alpha}g_{m\beta}g_{l\gamma}e_{\alpha\beta\gamma}U_{k,l}[\Phi - V(q)]d\Sigma -$$

$$- \sum_{q=1}^{Q_F}\iint_{\Sigma(q)} N_m(q)g_{m\alpha}g_{n\beta}\varepsilon^S_{\alpha\beta}\Phi_{,n}[\Phi - V(q)]d\Sigma \quad , \quad (1)$$

where $U_i$ is the *i*-th component of the elastic displacement vector, $\Phi$ is the piezoelectric potential (assuming their $\exp(j\omega t)$ time dependence), $N_m(p)$ and $N_m(q)$ are the *m*-th components of the unit vectors normal to the *p*-th exciting and *q*-th free electrodes, respectively, $E_{Gm}$ is the *m*-th component of the electric field strength generated within the local volume $\Omega_E$ of the plate by the applied signal, implying the relation $\vec{E}(x,y,z) = -\vec{\nabla}\Phi(x,y,z)$ holds, $c^E_{\alpha\beta\gamma\delta}$ are the components of the elastic tensor taken at a constant electric field strength, $e_{\alpha\beta\gamma}$ are the components of the piezoelectric tensor, $\varepsilon^S_{\alpha\beta}$ are the components of the dielectric tensor taken at a constant value of the elastic strain. The subscripts in Eq. (1) vary from 1 to 3 and the Einstein summation convention applies to repeated indices. The coma in the subscript marks differentiation. In Eq. (1), the constituent relations for components of the mechanical stress tensor and of the electric displacement vector are assumed to be linear in $U_i$ and $E_j = -\Phi_{,j}$.

Analyzing free vibrations, one should put $\Omega_E = 0$, $\sum_{p=1}^{Q_E} S(p) = 0$, $V_G(p) = 0$ and $\vec{E}_G = 0$. Furthermore, only the $E_G$ and $V_G(p)$ components written in Eq. (1) are related to forced vibrations. Therefore, when the exciting electrodes shown in Fig. 1(a) are all arranged on a single face of the plate the $\vec{E}_G$-components vanishes and $F$ reduces to the form given by EerNisse and Holland.[5] This allows to conclude that the vibration amplitude derived in the case of a surface excitation (when $E_{Gm} = 0$) is much less than the one attained with a bulk excitation (when $E_{Gm} \neq 0$).

Equating the first variation of $F$ to zero gives a standard set of equations in terms of four independent physical functions - $U_X(x,y,z)$, $U_Y(x,y,z)$, $U_Z(x,y,z)$, and $\Phi(x,y,z)$. It includes the equation of motion for extensional vibration, and the Maxwell's equation requiring that the divergence of the electric displacement vector to be zero. These should be supplemented with the boundary conditions on the plate boundary. On the dielectric part of the boundary, there are no normal components of both the mechanical stress and the electric displacement. On the part of the plate boundary covered with electrodes, the metal layers are considered to be thin and do not influence the above boundary conditions for the stress tensor.

To account for an arbitrary cut resonator plate, the functional (1) should be generalized by introducing the transition matrix $[g]$, which describes the coordinate transformation from the crystallographic system $\{X_C; Y_C; Z_C\}$ to the one $\{X; Y; Z\}$ defined by the plate cut. Then the matrix components $g_{\alpha\beta}$ are[24]

$$[g] = \begin{pmatrix} g_{11} & g_{12} & g_{13} \\ g_{21} & g_{22} & g_{23} \\ g_{31} & g_{32} & g_{33} \end{pmatrix} =$$
$$= \begin{pmatrix} \cos\alpha_2 \cos\alpha_3 & -\cos\alpha_2 \sin\alpha_3 & \sin\alpha_2 \\ \sin\alpha_1 \sin\alpha_2 \cos\alpha_3 + \cos\alpha_1 \sin\alpha_3 & -\sin\alpha_1 \sin\alpha_2 \sin\alpha_3 + \cos\alpha_1 \cos\alpha_3 & -\sin\alpha_1 \cos\alpha_2 \\ -\cos\alpha_1 \sin\alpha_2 \cos\alpha_3 + \sin\alpha_1 \sin\alpha_3 & \cos\alpha_1 \sin\alpha_2 \sin\alpha_3 + \sin\alpha_1 \cos\alpha_3 & \cos\alpha_1 \cos\alpha_2 \end{pmatrix}, \quad (2)$$

where $\alpha_1$, $\alpha_2$, and $\alpha_3$ are the rotation angles around the $X_C$, $Y_C$ and $Z_C$ axes, respectively.

With the electrode configuration shown in Fig 1(b), where the top and the bottom arrangements are mirror-symmetric to each other, the sum of the fifth, sixth and seventh terms in Eq. (1) reduces to

$$-\iiint_{\Omega_E} g_{m\alpha}g_{k\beta}g_{l\gamma}e_{\alpha\beta\gamma}E_{Gm}U_{k,l}d\Omega + \frac{1}{2}\iiint_{\Omega_E} g_{m\alpha}g_{n\beta}\varepsilon^S_{\alpha\beta}E_{Gm}\Phi_{,n}d\Omega -$$

$$-\frac{1}{2}\iiint_{\Omega_E} g_{m\alpha}g_{n\beta}\varepsilon^S_{\alpha\beta}E_{Gm}E_{Gn}d\Omega =$$

$$= -\sum_{r=1}^{Q_E/2}[\frac{V_G(r)}{L_y}]\iiint_{\Omega_E(r)} g_{m2}g_{k\beta}g_{l\gamma}e_{2\beta\gamma}U_{k,l}d\Omega +$$

$$+\frac{1}{2}\sum_{r=1}^{Q_E/2}[\frac{V_G(r)}{L_y}]\iiint_{\Omega_E(r)}(g_{2\alpha}g_{n\beta}\varepsilon^S_{\alpha\beta}\Phi_{,n} + g_{m\alpha}g_{2\beta}\varepsilon^S_{\alpha\beta}\Phi_{,m})d\Omega - \frac{1}{2}\sum_{r=1}^{Q_E/2}[\frac{V_G^2(r)}{L_y^2}]\iiint_{\Omega_E(r)} g_{2\alpha}g_{2\beta}\varepsilon^S_{\alpha\beta}d\Omega$$

Based on the above formulation, the free vibrations of a 128°-Y-rotated LiNbO$_3$ rectangular plate with the rotation angles $\alpha_1 = 128°$, $\alpha_2 = 0°$ and $\alpha_3 = 0°$ are then analyzed in three cases which are as follows (see Fig. 2).

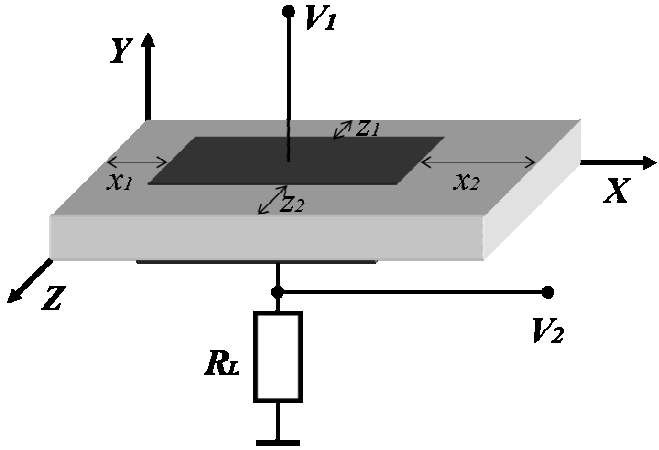

**Fig. 2.** Schematics of the impedance setup showing a LiNbO$_3$ plate (shaded body) oriented along the X- , Y- and Z- axes, metal electrodes (filled rectangles), the load resistance $R_L$, the input $V_1$ and output $V_2$ voltages.

(1) The plate surfaces are uncovered with metal electrodes, so $x_1 = x_2 = L_X/2$ and $z_1 = z_2 = L_Z/2$,

(2) A pair of open-circuited electrodes fully covers the opposite XZ-planes of the plate, so $x_1 = x_2 = 0$ and $z_1 = z_2 = 0$.

(3) A pair of open-circuited electrodes partly covers the opposite XZ-planes of the plate such that $x_1 = x_2 < L_X/2$ and $z_1 = z_2 = 0$.

Within the Rayleigh-Ritz calculation procedure, in order to minimize the $F$ one should expand the $U_X(x,y,z)$, $U_Y(x,y,z)$, $U_Z(x,y,z)$, and $\Phi(x,y,z)$ functions in series over a set of trial functions which must be complete and linear independent, and also continuous and doubly differentiable[25]. To this end the $BF_k(\alpha,x,y,z)$ trial functions with k = 1, 2, 3 and 4 corresponding to $U_X$, $U_Y$, $U_Z$ and $\Phi$, respectively, have been constructed in the following form

$$U_X(x,y,z) = \sum_{\alpha=1}^{Q_U} B(\alpha)\cdot BF_1(\alpha,x,y,z) =$$

$$= \left(\sum_{\alpha=1}^{Q_{U1}} B(\alpha)\cdot TF_{11}(\alpha,x,y,z) + \sum_{\alpha=Q_{U1}+1}^{Q_U} B(\alpha)\cdot TF_{12}(\alpha,x,y,z)\right)\cdot WF(K_X,K_Z,x,z) \qquad (3.1)$$

$$U_Y(x,y,z) = \sum_{\alpha=Q_U+1}^{2Q_U} B(\alpha)\cdot BF_2(\alpha,x,y,z) =$$

$$= \left(\sum_{\alpha=Q_U+1}^{Q_{U2}} B(\alpha)\cdot TF_{21}(\alpha,x,y,z) + \sum_{\alpha=Q_{U2}+1}^{2Q_U} B(\alpha)\cdot TF_{22}(\alpha,x,y,z)\right)\cdot WF(K_X,K_Z,x,z) \qquad (3.2)$$

$$U_Z(x,y,z) = \sum_{\alpha=2Q_U+1}^{3Q_U} B(\alpha)\cdot BF_3(\alpha,x,y,z) =$$

$$= \left(\sum_{\alpha=2Q_U+1}^{Q_{U3}} B(\alpha)\cdot TF_{31}(\alpha,x,y,z) + \sum_{\alpha=Q_{U3}+1}^{3Q_U} B(\alpha)\cdot TF_{32}(\alpha,x,y,z)\right)\cdot WF(K_X,K_Z,x,z) \qquad (3.3)$$

$$\Phi(x,y,z) = \sum_{\beta=1}^{Q_\Phi} C(\beta)\cdot BF_4(\beta,x,y,z) =$$

$$= \left(\sum_{\beta=1}^{Q_{\Phi 1}} C(\beta)\cdot TF_{41}(\beta,x,y,z) + \sum_{\beta=Q_{\Phi 1}+1}^{Q_{\Phi 2}} C(\beta)\cdot TF_{42}(\beta,x,y,z) + \sum_{\beta=Q_{\Phi 2}+1}^{Q_\Phi} C(\beta)\cdot TF_{43}(\beta,x,y,z)\right)\cdot WF(K_X,K_Z,x,z) \qquad .(3.4)$$

In Eqs.(3.1)-(3.3), the trigonometric trial functions $TF_{ij}(x,y,z)$ (i = 1, 2, 3) are taken from Table II in Ref. 4. It is convenient to approximate the potential $\Phi$ with

$$TF_{41}(\beta,x,y,z) = \sin\left(\frac{\pi M_{4S}(\beta)}{L_X}x\right)\cdot\cos\left(\frac{\pi N_{4C}(\beta)}{L_Y}y\right)\cdot\cos\left(\frac{\pi P_{4C}(\beta)}{L_Z}z\right), \qquad (4.1)$$

$$TF_{42}(\beta,x,y,z) = \cos\left(\frac{\pi M_{4C}(\beta)}{L_X}x\right)\cdot\sin\left(\frac{\pi N_{4S}(\beta)}{L_Y}y\right)\cdot\cos\left(\frac{\pi P_{4C}(\beta)}{L_Z}z\right), \qquad (4.2)$$

$$TF_{43}(\beta,x,y,z) = \sin\left(\frac{\pi M_{4S}(\beta)}{L_X}x\right) \cdot \sin\left(\frac{\pi N_{4S}(\beta)}{L_Y}y\right) \cdot \sin\left(\frac{\pi P_{4S}(\beta)}{L_Z}z\right), \qquad (4.3)$$

Here, the integers $M_{kS}(\alpha)$, $M_{kC}(\alpha)$, $N_{kS}(\alpha)$, $N_{kC}(\alpha)$, $P_{kS}(\alpha)$ and $P_{kC}(\alpha)$ are the coordinate characteristic numbers (CCNs), and the number of the trial functions is defined by $Q_U$ and $Q_\Phi$. Following the landmark paper,[4] the integers in the $\{M_{kS}(\alpha), M_{kC}(\alpha)\}$-, $\{N_{kS}(\alpha), N_{kC}(\alpha)\}$- and $\{P_{kS}(\alpha), P_{kC}(\alpha)\}$- pairs are chosen to be of the opposite parities, namely, {odd, even} or {even, odd}. The $B(\alpha)$ and $C(\beta)$ coefficients are estimated by a standard procedure.[5]

Apparently, there are two important issues in constructing the trial functions needed for characterizing the vibrations of resonators which are partly covered by electrodes. First, a new view emerges from the terms $TF_{41}(x,y,z)$ and $TF_{42}(x,y,z)$ which are introduced in Eq. (4) in order to account for the penetration of the electric field outside the vibrating plate. Previously, the expansion of the $\Phi$ function into series of two $TF_{4j}$ components including the $\sin(...)\cdot\sin(...)\cdot\sin(...)$ and $\cos(...)\cdot\cos(...)\cdot\sin(...)$ terms[4] was only capable to describe the vibrations of a short-circuited piezoelectric plate covered fully with electrodes.

Second, a proper understanding requires taking into account the wave-factor $WF(K_X, K_Y, x, z)$ included in Eq. (3). This comes from the plate waves traversing the plate which are likely to appear if the electrode area is remarkably small thus making a significant contribution to vibration properties of the plate.[17] Setting $WF(K_X, K_Z, x, z) = 1$ allows to attain a large-electrode area case.

For definiteness, suppose the waves are purely standing and assume their reflections at the edges of the plate are loss-free. Then

$$WF(K_X, K_Z, x, z) = WF_1(K_X, x) \cdot WF_2(K_Z, z), \qquad (5.1)$$

$$WF_1(K_X, x) = \frac{1}{2}\{\cos(K_X x) + \cos[K_X(L_X - x)]\}, \qquad (5.2)$$

$$WF_2(K_Z, z) = \frac{1}{2}\{\cos(K_Z z) + \cos[K_Z(L_Z - z)]\}. \qquad (5.3)$$

where $K_X = \omega/V_X(\omega)$ and $K_Z = \omega/V_Z(\omega)$ are the X- and Z- components of the wave vector, respectively, and $V_X$ and $V_Z$ are the corresponding components of the wave phase velocity.

Turning to the trial functions introduced here it should be noted that the sets of $\{TF_{11}, TF_{12}\}$, $\{TF_{21}, TF_{22}\}$, $\{TF_{31}, TF_{32}\}$, and $\{TF_{41}, TF_{42}, TF_{43}\}$ are neither orthogonal nor complete, although, taken separately, each of the sets $\{TF_{ij}\}$ definitely satisfies the general demand of the orthogonality and completeness. As remarked previously, such "overcomplete" sets of the trial functions accelerate the convergence of the series (3) in computations.[26] Meanwhile, the mathematics behind this problem is well beyond the scope of the present work. Concerning the wave-factor WF given above, it is easy to show that, for $K_X L_X = \pi N_X$ or $K_Z L_Z = \pi N_Z$ ($N_X, N_Z = 1, 2, ...$), which are only of practical relevance since in these cases the surface distributions of the elastic and electric fields are remarkably affected by the $WF(K_X, K_Y, x, z)$ wave factor, the $\{BF_{ij}\}$ trial function sets remain complete and orthogonal. Therefore, the computation results require further experimental evidence to secure the validity of the trial functions introduced above. As shown below, the principal findings of experimental vibration frequencies match the expectations of this theory thus supporting the theoretical approach.

When treating the vibration problem variationally for a particular vibration mode, the distributions of $U_X$, $U_Y$, $U_Z$ and $\Phi$ over the surface of a vibrating plate are defined by a proper choice of the odd ("O") and even ("E") parities, which are assigned to the corresponding CCN.[4] Taking $U_Z$ component as an example, this can be written as

$$U_Z(L_X, y, z) = U_Z(0, y, z) \text{ for the even } M_Z(\alpha),$$

$$U_Z(L_X, y, z) = -U_Z(0, y, z) \text{ for the odd } M_Z(\alpha)$$

due to the presence of the *cos*-components in the expansion of $U_Z$ given in Table II of Ref. 4. These formulas indicate that the asymmetric and symmetric $U_Z(X)$ distributions occur. Speaking more generally, there would be eight parity sets $\{R1; R2; R3; R4; R5; R6; R7; R8\}$ for the CCN triplet of $\{M_k(\alpha), N_k(\alpha), P_k(\alpha)\}$ in the trial function $TF_{ij}(x, y, z)$. Then $R1 = (E,E,E)$, $R2 = (O,E,E)$, $R3 = (E,O,E)$, $R4 = (E,E,O)$, $R5 = (E,O,O)$, $R6 = (O,E,O)$, $R7 = (O,O,E)$ and $R8 = (O,O,O)$, where the first parity is ascribed to the X-coordinate CCN, the second parity – to that of the Y-coordinate, and the third parity – to that of the Z-coordinate.

Therefore, to model the plate vibrations, taking into account the opposite parities of the CCN pairs, four parity sets *Rn* corresponding to twelve CCNs in Eqs. (3) should be specified for every vibration mode. As far as we are aware, this has not yet been done in the 3D-resonators.

In general, the total number of the twelve parity combinations is $2^{12} = 4096$, corresponding to the number of extensional vibration modes which are likely to occur in the rectangular piezoelectric parallelepiped. However, the analysis done allows to conclude that, in the absence of the mechanical load and free charges on the dielectric part of the plate boundary, all these vibration modes can be subdivided into 512 groups, 8 modes in each group. One of them applicable to the piezoceramic parallelepiped was described by Holland.[4] Within the modal group, a certain set of material constants (MCs), $c^E_{\alpha\beta\gamma\delta}$, $e_{\alpha\beta\gamma}$ and $\varepsilon^S_{\alpha\beta}$, is used in the computations, independently of the way of distributing the various trigonometric trial functions over $U_X$, $U_Y$, $U_Z$ and $\Phi$. Consequently, every mode in a given group has the same eigenfrequency spectrum reducing the total number of different MCs to much less than 512. Therefore, it is a few different vibration modes that are only sustained in the plate of a given crystallographic class.

Any theory addressing the coupled electroelastic vibrations must explain one general finding. Depending on the plate cut, some of the vibration modes are not piezoelectrically active and, therefore, can not be sustained by electrically exciting the plate. In order to check the consistency of our theoretical model, a pure cut LiNbO$_3$ plate has been tested. It appears that there are no $e_{\alpha\beta\gamma} \neq 0$ entries in some of the 512 MCs, consistent with the above general demand.

A wide range of solutions to forced vibration problems is available with this technique. For definiteness, in the discussion that follows two questions are answered.

1. Which vibration modes can be excited with a given electrode configuration?

2. Which electrode configuration and distribution of the input signals over the electrodes allow to support a given vibration mode?

The fifth and the sixth terms in the functional (1) are the key to answering these questions. In general the electrical excitation of the vibrations is possible if the two terms are nonzero. It might be expected that the number of modes supported by the plate takes its minimum value using the $\vec{E}_G$ exciting field with only one nonzero component. This case can be realized, for example, in the geometry shown in Fig. 1(b) if the inter-electrode spacing is large enough to give

$$\iiint_{\Omega_E} g_{m\alpha} g_{k\beta} g_{l\gamma} e_{\alpha\beta\gamma} U_{k,l} d\Omega \neq 0 \qquad (6)$$

for $TF_{1i}$, $TF_{2i}$ and $TF_{3i}$ (Table II in Ref. 4) with $\alpha = 1$, 2 and 3 for *YZ*-, *XZ*- and *XY*-faces, respectively; and

$$\iiint_{\Omega_E} (g_{k\alpha} g_{n\beta} \varepsilon^S_{\alpha\beta} \Phi_{,n} + g_{m\alpha} g_{k\beta} \varepsilon^S_{\alpha\beta} \Phi_{,m}) d\Omega \neq 0 \qquad (7)$$

for $TF_{4j}$ [see Eq. (4)] with $k = 1$, 2 and 3 for *YZ*-, *XZ*-, and *XY*-faces, respectively. Noticeably, relations (6) and (7) are applicable to a piezocrystal plate of any crystallographic class, standing as a criterion mediating between the electrically-excited and non-excited vibration modes.

Substituting the trial functions modeling $U_i$ and $\Phi$ into (6) and (7) and integrating over $\Omega_E$, which is defined by the electrode configuration, one takes the parity sets related to the excited vibration modes. Alternatively, the parity sets forcing the integrals in formulas (6) and (7) to be zero define non-piezoelectric vibration modes which can therefore not be electrically excited. This analysis allows to correctly choose the parity sets pertinent to the measured frequency spectrum, but requires further experimental evidence to answer the question if the mode is non-degenerate or not.

Some of the computation results are summarized in Table I. Given here are the parity sets *Rn* for the vibration modes of a 128°-Y-rotated cut LiNbO$_3$ rectangular plate which is electrically excited with a single pair of rectangular-shaped electrodes (see Fig. 2). To determine the parity sets substitute $\alpha = 1$ and $k = 2$ in formulas (6) and (7), respectively, so the integration reduces to

$$\iiint_{\Omega_E} (...) d\Omega \rightarrow \int_{x1}^{L_X - x2} dx \int_0^{L_Y} dy \int_{z1}^{L_Z - z2} (...) dz .$$

**TABLE I**. The parity sets $Rn$ for the trial functions $TF0_{ij}$ used to model the extensional vibration modes electrically excited in a 128°-Y-rotated $LiNbO_3$ rectangular-plate resonator with a pair of rectangular-shaped electrodes located asymmetrically (A) or symmetrically (B) on its opposite faces. Electrode's dimensions are given in the first column.

|  | For $TF0_{1i}$ | For $TF0_{2i}$ | For $TF0_{3i}$ | For $TF0_{4j}$ |  |
|---|---|---|---|---|---|
| (A) Asymmetric location ||||||
| (A1) Electrodes are on *XY*-faces ||||||
| $x_1=x_2=0, y_1=y_2=0$ | R5 | R6 | R7 | R5, R7 | 2 |
| $x_1 \neq x_2, y_1=y_2=0$ | R5, R8 | R4, R6 | R3, R7 | R3, R5, R7, R8 | 32 |
| $x_1=x_2=0, y_1 \neq y_2$ | R4, R5 | R6, R8 | R2, R7 | R2, R4, R5, R7 | 32 |
| $x_1 \neq x_2, y_1 \neq y_2$ | R4, R5, R6, R8 | R4, R5, R6, R8 | R2, R3, R7, R8 | R1, R2, R3, R4, R5, R6, R7, R8 | 512 |
| (A2) Electrodes are on *XZ*-faces ||||||
| $x_1=x_2=0, z_1=z_2=0$ | R5 | R6 | R7 | R6 | 1 |
| $x_1 \neq x_2, z_1=z_2=0$ | R5, R8 | R4, R6 | R3, R7 | R4, R6 | 16 |
| $x_1=x_2=0, z_1 \neq z_2$ | R3, R5 | R2, R6 | R7, R8 | R2, R6 | 16 |
| $x_1=x_2=0, z_1=z_2=0$ | R5 | R6 | R7 | R6 | 1 |
| (A3) Electrodes are on *YZ*-faces ||||||
| $y_1=y_2=0, z_1=z_2=0$ | R5 | R6 | R7 | R5, R7 | 2 |
| $y_1 \neq y_2, z_1=z_2=0$ | R4, R5 | R6, R8 | R2, R7 | R2, R4, R5, R7 | 32 |
| $y_1=y_2=0, z_1 \neq z_2$ | R3, R5 | R2, R6 | R7, R8 | R3, R5, R7, R8 | 32 |
| $y_1 \neq y_2, z_1 \neq z_2$ | R1, R2, R4, R5 | R2, R6, R7, R8 | R2, R6, R7, R8 | R1, R2, R3, R4, R5, R6, R7, R8 | 512 |
| (B) Symmetric location ||||||
| (B1) Electrodes are on *XY*-faces ||||||
| $x_1 = x_2 \neq 0, y_1=y_2=0$ | R5 | R6 | R7 | R5, R7 | 2 |
| $x_1=x_2=0, y_1 = y_2 \neq 0$ | R5 | R6 | R7 | R5, R7 | 2 |
| $x_1 = x_2 \neq 0, y_1 = y_2 \neq 0$ | R5 | R6 | R7 | R5, R7 | 2 |
| (B2) Electrodes are on *XZ*-faces | | | | | |
| $x_1 = x_2 \neq 0, z_1 = z_2 = 0$ | R5 | R6 | R7 | R6 | 1 |
| $x_1 = x_2 = 0, z_1 = z_2 \neq 0$ | R5 | R6 | R7 | R6 | 1 |
| $x_1 = x_2 \neq 0, z_1 = z_2 \neq 0$ | R5 | R6 | R7 | R6 | 1 |
| (B3) Electrodes are on *YZ*-faces | | | | | |
| $y_1 = y_2 \neq 0, z_1 = z_2 = 0$ | R5 | R6 | R7 | R5, R7 | 2 |
| $y_1 = y_2 = 0, z_1 = z_2 \neq 0$ | R5 | R6 | R7 | R5, R7 | 2 |
| $y_1 = y_2 \neq 0, z_1 = z_2 \neq 0$ | R5 | R6 | R7 | R5, R7 | 2 |

All the *Rn* sets given in Table I correspond to the trial function of the form

$$TF0_{ki}(x,y,z) = \sin[\pi m_{ki}(\alpha) x / L_X] \cdot \sin[\pi n_{ki}(\alpha) y / L_Y] \cdot \sin[\pi p_{ki}(\alpha) / L_Z]$$

with $i = 1$ and 2 for $k = 1, 2$ and 3, and $i = 1, 2$ and 3 for $k = 4$, since the $\{M_{kS}(\alpha), M_{kC}(\alpha)\}$-, $\{N_{kS}(\alpha), N_{kC}(\alpha)\}$- and $\{P_{kS}(\alpha), P_{kC}(\alpha)\}$-pairs are of the opposite parities. In terms of the parity sets $\{R1; R2; R3; R4; R5; R6; R7; R8\}$, each extensional vibration mode can be described by

$$\{Ri(\text{for } TF0_1), Rj(\text{for } TF0_2), Rk(\text{for } TF0_3), Rl(\text{for } TF0_4)\} \equiv \{Ri, Rj, Rk, Rl\} \ (i,j,k,l = \overline{1,8}).$$

Then the case of $x_1 = x_2$ and $z_1 = z_2$ in Fig.2 is referred to as the symmetric electrode configuration. Otherwise, it is non-symmetric. In case of the multiple pairs of electrodes shown in Fig.1(a) the electrode configuration is symmetric if the bottom- and top-faced electrodes are mirrored with respect to the middle plane of the plate.

Obviously, the non-symmetric electrode configurations produce a greater number of vibration modes due to a greater number of the nonzero $E_G$ components; see Eq. (1). The number of various vibration modes supported by a given electrode configuration is summarized in the last column of Table I. However, there is one significant issue that will need further work: It is still unclear which of the modes are non-degenerated, giving rise to different eigenfrequency spectra. Apparently, a better understanding requires taking into account several different vibration modes that can be excited simultaneously at a given frequency. In this case, the elastic displacements and the piezoelectric potential can be expanded over the normal vibration modes as illustrated, e.g., in Refs. 5 and 27.

## III. SAMPLES

Experiments were performed on two samples, LNO-1 and LNO-4a, of a $128^o$-Y-rotated cut lithium niobate ($\{\alpha_1,\alpha_2,\alpha_3\}=\{128^o,0^o,0^o\}$) which is typically used for guiding the surface acoustic waves[28]. Experimentally, have been investigated. The samples were rectangular shaped with linear dimensions $L_X$, $L_Y$ and $L_Z$ given in Table II. Also tabulated is a measure of the nonuniformity in $L_i$ defined as $\Delta L_i = L_{i,max} - L_{i,min}$ with $i = x, y, z$.

**TABLE II.** Sample's average linear dimensions ($\overline{L}_i$) and related nonuniformities ($\Delta L_i$) (in mm).

| Sample | $\overline{L}_x$ | $\Delta L_x$ | $\overline{L}_y$ | $\Delta L_y$ | $\overline{L}_z$ | $\Delta L_z$ |
|---|---|---|---|---|---|---|
| LNO-1 | 13.45 | 0.01 | 0.755 | 0.005 | 3.64 | 0.01 |
| LNO-4a | 19.20 | 0.02 | 0.70 | 0.01 | 3.61 | 0.01 |

Electroelastic vibrations were excited with a pair of thin metallic electrodes formed by a fine-dispersive Ag powder deposited on the opposite XZ-planes of the samples. The tangent of the dielectric loss, measured at the frequency of 1 kHz, did not exceed $5\times10^{-3}$. Care was taken to ensure the stability of the electrical input and output contacts attached to the electrodes and to achieve the traction-free mechanical condition at the surfaces of the plate.

## IV. EIGENFREQUENCY MEASUREMENTS

The eigenfrequencies of the samples were taken using the impedance technique[29] schematically sketched in Fig. 2. Taking the load resistance $R_L$ to be much greater than the real part $\mathrm{Re}\{\dot{Z}(f)\}$ of the sample's electric impedance $\dot{Z}(f)$ this gives

$$\left|\dot{Y}(f)\right| \approx R_L^{-1} \cdot \left|\dot{V}_2(f)/\dot{V}_1(f)\right|,$$

where $\dot{Y}(f) = \dot{Z}^{-1}(f)$ is the complex electrical conductivity of the sample and the voltages $\dot{V}_1(f)$ and $\dot{V}_2(f)$ are shown in Fig. 2. Then the frequency dependence $\left|\dot{Y}(f)\right|$ represents the sample's eigenfrequency spectrum.[30] This was found to change with varying the values of $x_1$, $x_2$, $z_1$, and $z_2$.

## V. EXPERIMENTAL RESULTS AND DISCUSSION

The frequency dependence $\left|\dot{Y}(f)\right|$ taken in sample LNO-1 fully covered with electrodes (i.e. $x_1 = x_2 = 0$ and $z_1 = z_2 = 0$ in Fig. 2) is exemplified in Fig. 3. A number of maxima and minima reflecting an overtone's family of a vibration mode (or, maybe, of several modes) is clearly resolved.

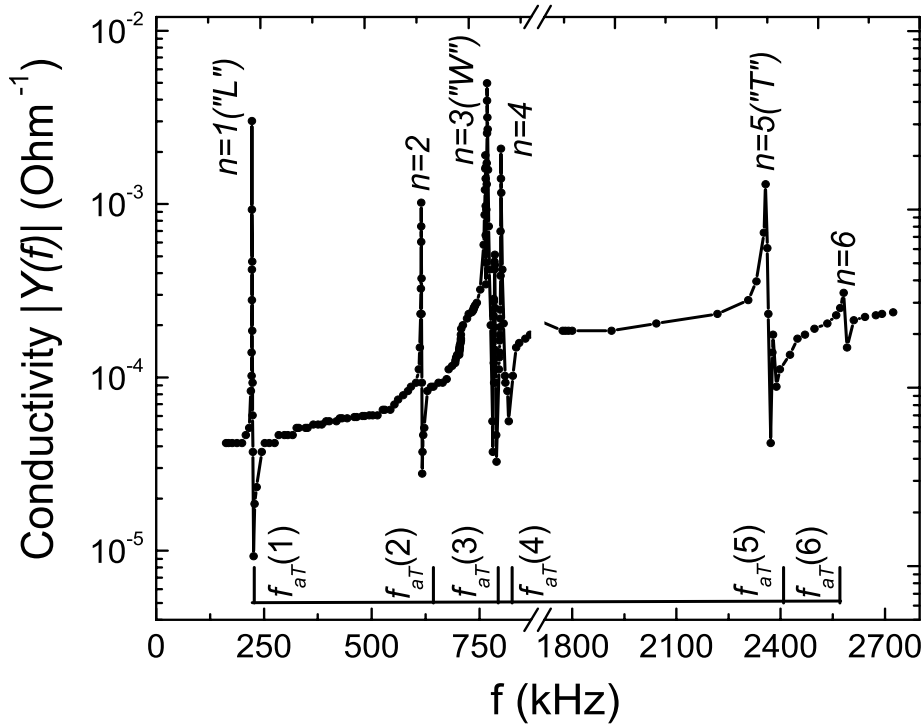

**Fig. 3.** Frequency dependence of the complex conductivity $\left|\dot{Y}(f)\right|$ taken in sample LNO-1 (solid curve). The computed antiresonant frequencies $f_{aT}(n)$ are martked below the curve.

The measured frequencies $f_{r,E}(n)$ corresponding to maxima in Fig. 3 are the parallel-resonance frequencies of the sample with $n$ being the serial number of an overtone. Minima $f_{a,E}(n)$ give the sample's series-resonance (or antiresonant) frequencies. It is known[31] that the eigenfrequency spectrum of a piezoelectric resonant structure is represented by either the $f_r(n)$ set for the shirt-circuited electrodes or the $f_a(n)$ set for the open-circuited ones. Theoretically, the setup displayed in Fig. 2 is considered to be open-circuited. The most significant resonant and antiresonant frequencies taken in our samples are summarized in Table III.

**TABLE III**. A comparison between the experimentally measured frequencies $f_{a,E}(n)$ and the computed frequencies $f_{aT}(n)$ for LNO-1 and LNO-4a samples. The frequencies $f_{aT}(n)$ are calculated with $Q_U = Q_\Phi = 33$ for the {R5,R6,R7,R6}-vibration mode.

| | Overtone's serial number $n$ | | | | | |
|---|---|---|---|---|---|---|
| | $n=1$ ("L") | $n=2$ | $n=3$ ("W") | $n=4$ | $n=5$ ("T") | $n=6$ |
| Sample LNO-1 | | | | | | |
| $f_{r,E}(n)$, kHz | 222.2 | 614.8 | 768.0 | 799.9 | 2357.1 | 2581.6 |
| $f_{a,E}(n)$, kHz | 226.1 | 617.3 | 781.1 | 830.8 | 2371.2 | 2591.0 |
| $f_{a,T}(n)$, kHz | 227.3 | 643.0 | 793.4 | 826.2 | 2408.7 | 2570.5 |
| $\Delta_1(n)$, % | 0.5 | 3.9 | 1.5 | 0.6 | 1.6 | 0.8 |
| Sample LNO-4a | | | | | | |
| $f_{r,E}(n)$, kHz | 158.3 | 461.5 | 731.6 | 764.1 | 2259.8 | 2492.8 |
| $f_{a,E}(n)$, kHz | 160.3 | 462.0 | 738.9 | 782.4 | 2291.1 | 2503.7 |
| $f_{a,T}(n)$, kHz | 159.5 | 472.4 | 723.3 | 795.9 | 2327.6 | 2562.6 |
| $\Delta_1(n)$, % | 0.5 | 2.2 | 2.1 | 1.7 | 1.6 | 2.3 |

In order to compare the measured and computed frequencies, suppose that the twist-like vibrations are not excited with the electrode configuration shown in Fig. 2 and the observed vibrations are pure flexural and extensional ones or coupling of the two. Starting with the one-dimensional case, the fundamental frequencies of the length-propagating ($f_a^L$), thickness-propagating ($f_a^T$) and width-propagating ($f_a^W$) vibration modes marked here as "L", "T" and "W", respectively, can be approximated by[29]

$$f_a^L = V_X / 2\bar{L}_X, \quad f_a^T = V_Y / 2\bar{L}_Y, \quad f_a^W = V_Z / 2\bar{L}_Z. \tag{8}$$

Taking the phase velocities $V_X \approx 6500$ m/s for a *longitudinal* wave, $V_Y \approx 4000$ m/s for a *fast shear* wave[32] and the values of $\bar{L}_X$, $\bar{L}_Y$ for sample LNO1 in Table II, the estimates are $f_a^L \approx 242$ kHz and $f_a^T \approx 2649$ kHz. Comparison with the data shown in Table IIIA exhibits pure agreement of the measured and computed values since, experimentally, $f_{aE}(1) \approx 226$ kHz and $f_{aE}(5) \approx 2371$ kHz. Taking $V_Z \approx 6500$ m/s for a *longitudinal* wave[32] and the value of $\bar{L}_Z$ in Table II yields $f_a^W \approx 892$ kHz which, again, is remarkably greater than the measured value $f_{aE}(3) \approx 781$ kHz. Therefore, a simple one-dimensional model obviously overestimates the vibration frequencies.

The other $f_{aE}$ frequencies observed experimentally and given in Table III are also inconsistent with the estimated values obtained within a simplified three-dimensional standing-wave approach.[18] Therefore, it can be suggested that the peaks seen in Fig. 3 are the overtones of the 3D extensional vibration modes.

This way of looking at the experimental results can be given supporting evidence by using the computation work performed here. The computed $f_{aT}(n)$ and the measured $f_{rE}(n)$ and $f_{aE}(n)$ frequencies are contrasted in Table III. In the plate with fully metallized XZ-planes, the {*R5,R6,R7,R6*}-vibration mode should be analyzed; see second row [part (B2)] in Table I. Computing the eigenfrequencies, the material constants $c_{ij}^E$, $e_{mn}$ and $\varepsilon_{kl}^S$ are taken from.[33] The spectrum $f_{aT}(n)$ is obtained with $Q_U = Q_F = 33$. Following the classification scheme given above the *R*8 and *R*2 parity sets are used for the $TF_{11}$ and $TF_{12}$ trial functions, *R*8 and *R*3 – for $TF_{21}$ and $TF_{22}$, *R*8 and *R*4 – for $TF_{31}$ and $TF_{32}$, *R*7, *R*1 and *R*6 – for $TF_{41}$, $TF_{42}$ and $TF_{43}$, respectively.

Comparing the measured and computed eigenfrequencies given in Table III, a few percent discrepancy is seen which can be due to two main factors. First one comes from the non-uniformity of the sample dimensions (see $\Delta L_X$, $\Delta L_Y$ and $\Delta L_Z$ in Table II). Evidently, decrease in $L_i$ across the plate increases $f_{aT}(n)$. For example, a 1% decrease (increase) in $L_X$, $L_Y$ and $L_Z$ yields the frequency $f_{aT}(1)$ increase (decrease) of about 1% in sample LNO-1. Also, the non-uniformity produces a departure $\Delta\alpha_i$ from the angles $\alpha_i$ defining the crystallographic cut of the sample. Taking $\Delta\alpha_1 \approx 1^o$, $\Delta\alpha_2 \approx 1^o$, $\Delta\alpha_3 \approx 1^o$ in sample LNO-1 gives the $f_{aT}(1)$ frequency shift from 227.3 to 227.2 kHz for $\{\alpha_1,\alpha_2,\alpha_3\}=\{129^o,1^o,1^o\}$. If $\{\alpha_1,\alpha_2,\alpha_3\}=\{127^o,-1^o,-1^o\}$ the frequency shifts from 227.3 to 227.4 kHz. Second, the values of the material constants are not definitely known. Then, varying $c_{ij}^E$, $e_{mn}$ and $\varepsilon_{kl}^S$ within ±1% gives the $f_{aT}(1)$ shift of about ±0.5% in sample LNO-1. Therefore, a comparison of the observed frequency spectrum with variational calculations, using the functional and the trial functions given above, allows to identify the majority of the peaks shown in Fig. 3. However, there exist fine structures in the frequency spectra that can still not be explained. For example, the peak marked by "W" in Fig.3 exhibits the triplet structure: It is accompanied by two subsidiary peaks on its low- and high-frequency side. The explanation may lie in additional effects, such as the coupling between the tone and flexural vibration modes, which are beyond the scope of the present work.

Experiments show the significant systematic effect of decreasing eigenfrequencies with increasing the metallization surface area, and this is supported by the computations. One example is shown in Table IV where the frequency spectra are given for two symmetric electrode configurations (see Fig. 2), taken both theoretically and experimentally.

**TABLE IV.** The experimentally measured frequencies $f_{a,E}(n)$, and the computed frequencies $f_{a,T}(n)$, for various values of electrode's area (sample LNO-1, symmetric configuration of electrodes).

| | Overtone's serial number $n$ | | | | | |
|---|---|---|---|---|---|---|
| | n=1 ("L") | n=2 | n=3 ("W") | n=4 | n=5 ("T") | n=6 |
| Calculated values $f_{a,T}(n)$ [kHz], ($Q_U=Q_\Phi=33$, the {R5,R6,R7,R6}-vibration mode) | | | | | | |
| Non-metallized faces | 229.6 | 683.1 | 822.9 | 877.1 | 2583.0 | 2667.5 |
| 70%-metallized | 227.8 | 652.7 | 804.9 | 843.0 | 2526.1 | 2625.3 |
| 100%-metallized | 227.3 | 643.0 | 793.4 | 826.2 | 2408.7 | 2570.5 |
| $\delta f_{aT}(n)$, % | 0.2 | 1.5 | 1.45 | 2.0 | 4.8 | 2.1 |
| Measured values $f_{a,E}(n)$, kHz | | | | | | |
| 70%-metallized | 227.1 | 620.6 | 788.9 | 838.5 | 2392.6 | 2607.0 |
| 100%-metallized | 226.1 | 617.3 | 781.1 | 830.8 | 2371.2 | 2591.0 |
| $\delta f_{aE}(n)$, % | 0.4 | 0.5 | 1.0 | 0.9 | 0.9 | 0.6 |

These are fully-metallized XZ-faces with $x_1 = x_2 = 0$ mm and $z_1 = z_2 = 0$ mm, and ≈ 70%-metallized XZ-faces with $x_1 = x_2 = 2$ mm and $z_1 = z_2 = 0$ mm. Here, the results are contrasted with that taken in the non-metallized sample. It is seen that, at any given overtone serial number $n$, metallization decreases both the $f_{aT}$ and $f_{aE}$ frequencies.

## VI. CONCLUSIONS

The following conclusions can be drawn from the presented analysis.

(1) The problem of the free and forced (electrically excited) extensional vibrations in a three-dimensional rectangular plate arbitrarily cut from a piezoelectric crystal can be solved variationally by using the generalized functional, quadratic in the elastic displacement components and the piezoelectric potential, proposed in this work.

(2) The vibration mode classification scheme applicable to a purely extensional vibration of piezoelectric parallelepipeds is developed.

(3) A good agreement between the measured and computed antiresonant frequencies in a 128°-Y-rotated cut LiNbO$_3$ rectangular plate is obtained for a few lowest overtones in the electrically excited vibration mode.

(4) The natural frequencies of the extensional vibration mode overtones are found to decrease, both experimentally and theoretically, with increasing the area of the exciting electrodes.

## ACKNOWLEDGMENTS

We are grateful to Eugen V. Psyarnetsky and Eugen O. Melezhyk for their assistance in making measurements.